\documentstyle[epsf]{mymn}

\newif\ifAMStwofonts

\ifoldfss
  \ifCUPmtlplainloaded \else
    \NewTextAlphabet{textbfit} {cmbxti10} {}
    \NewTextAlphabet{textbfss} {cmssbx10} {}
    \NewMathAlphabet{mathbfit} {cmbxti10} {} 
    \NewMathAlphabet{mathbfss} {cmssbx10} {} 
  \fi
  \ifAMStwofonts
    \ifCUPmtlplainloaded \else
      \NewSymbolFont{upmath} {eurm10}
      \NewSymbolFont{AMSa} {msam10}
      \NewMathSymbol{\upi}     {0}{upmath}{19}
      \NewMathSymbol{\umu}     {0}{upmath}{16}
      \NewMathSymbol{\upartial}{0}{upmath}{40}
      \NewMathSymbol{\leqslant}{3}{AMSa}{36}
      \NewMathSymbol{\geqslant}{3}{AMSa}{3E}

       \let\le=\leqslant
       \let\ge=\geqslant
    \fi
  \fi
\fi 

\ifnfssone
  \newmathalphabet{\mathit}
  \addtoversion{normal}{\mathit}{cmr}{m}{it}
  \addtoversion{bold}{\mathit}{cmr}{bx}{it}
  \newmathalphabet{\mathbfit} 
  \addtoversion{normal}{\mathbfit}{cmr}{bx}{it}
  \addtoversion{bold}{\mathbfit}{cmr}{bx}{it}
  \newmathalphabet{\mathbfss} 
  \addtoversion{normal}{\mathbfss}{cmss}{bx}{n}
  \addtoversion{bold}{\mathbfss}{cmss}{bx}{n}
  \ifAMStwofonts
    \ifCUPmtlplainloaded \else
      \UseAMStwoboldmath
      \makeatletter
      \new@mathgroup\upmath@group
      \define@mathgroup\mv@normal\upmath@group{eur}{m}{n}
      \define@mathgroup\mv@bold\upmath@group{eur}{b}{n}
      \edef\UPM{\hexnumber\upmath@group}
      \new@mathgroup\amsa@group
      \define@mathgroup\mv@normal\amsa@group{msa}{m}{n}
      \define@mathgroup\mv@bold\amsa@group{msa}{m}{n}
      \edef\AMSa{\hexnumber\amsa@group}
      \makeatother
      \mathchardef\upi="0\UPM19
      \mathchardef\umu="0\UPM16
      \mathchardef\upartial="0\UPM40
      \mathchardef\leqslant="3\AMSa36
      \mathchardef\geqslant="3\AMSa3E

       \let\le=\leqslant
       \let\ge=\geqslant
    \fi
  \fi
\fi 

\ifnfsstwo
  \DeclareMathAlphabet{\mathbfit}{OT1}{cmr}{bx}{it}
  \SetMathAlphabet\mathbfit{bold}{OT1}{cmr}{bx}{it}
  \DeclareMathAlphabet{\mathbfss}{OT1}{cmss}{bx}{n}
  \SetMathAlphabet\mathbfss{bold}{OT1}{cmss}{bx}{n}
  \ifAMStwofonts
    \ifCUPmtlplainloaded \else
      \DeclareSymbolFont{UPM}{U}{eur}{m}{n}
      \SetSymbolFont{UPM}{bold}{U}{eur}{b}{n}
      \DeclareSymbolFont{AMSa}{U}{msa}{m}{n}
      \DeclareMathSymbol{\upi}{0}{UPM}{"19}
      \DeclareMathSymbol{\umu}{0}{UPM}{"16}
      \DeclareMathSymbol{\upartial}{0}{UPM}{"40}
      \DeclareMathSymbol{\leqslant}{3}{AMSa}{"36}
      \DeclareMathSymbol{\geqslant}{3}{AMSa}{"3E}

       \let\le=\leqslant
       \let\ge=\geqslant
    \fi
  \fi
\fi 

\ifCUPmtlplainloaded \else
  \ifAMStwofonts \else 
    \def\upi{\pi}
    \def\umu{\mu}
    \def\upartial{\partial}
  \fi
\fi

\title[Galactic Classification with Decision Trees]{Using Oblique
Decision Trees for the Morphological Classification of Galaxies}
\author[E. A. Owens et al.]{
E. A. Owens, R. E. Griffiths, K. U. Ratnatunga \\
Bloomberg Center for Physics and Astronomy, Johns
Hopkins University, Homewood Campus, Baltimore, MD 21218,
USA
}

\pagerange{\pageref{firstpage}--\pageref{lastpage}}
\pubyear{1996}

\begin{document}

\maketitle
\label{firstpage}

\begin{abstract}

We discuss the application of a class of machine learning algorithms
known as decision trees to the process of galactic classification.  In
particular, we explore the application of oblique decision trees
induced with different impurity measures to the problem of classifying
galactic morphology data provided by Storrie-Lombardi et al.
\shortcite{st}.  Our results are compared to those obtained by a neural
network classifier created by Storrie-Lombardi et al, and we show that
the two methodologies are comparable.  We conclude with a
demonstration that the original data can be easily classified into
less well-defined categories.

\end{abstract}


\section{Introduction}

Decision tree algorithms have proven themselves useful as automated
classifiers in a number of astronomical domains.  An oblique decision
tree algorithm created by Murthy, Kasif, \& Salzberg (1994) has
demonstrated that decision trees can be generated to distinguish
between stars and galaxies or to identify cosmic rays in Space
Telescope images.  Typically, the trees produced by this algorithm
possess accuracies up to and exceeding 95\%, and can now be used to
classify additional data
\cite{sa2}.  In this way, decision trees free researchers
from the tedious task of object classification.

        The next logical step for a decision tree classification
algorithm is, of course, the classification of galaxies
morphologically.  As Storrie-Lombardi et al. point out \shortcite{st},
this has been attempted by a number of researchers with limited
success; today ``morphological classification into ellipticals,
lenticulars, spirals and irregulars remains a process dependent on the
eyes of a handful of dedicated individuals.''

In an attempt to rectify this situation, Storrie-Lombardi et al.
(hereafter referred to as SLSS) have applied a computing technique
known as artificial neural networks (also known as neural nets or
ANNs), to the problem of galaxy classification.  In particular, SLSS
used their neural net to classify galaxies taken from the ESO-LV
catalog \cite{lau}.  In this paper, we compare results from the SLSS
ANN with those from decision trees.

\section{Neural Nets and Decision Trees}

Here we will describe some of the fundamental differences between the
neural network algorithm implemented by SLSS and decision trees.  Our
discussion will focus only on the dominant aspects of these
algorithms.  For more information, the interested reader is referred
to the SLSS paper on neural network classification \cite{st}.  For an
introduction to decision trees, Quinlan's original paper
\shortcite{qu1} on the topic is an excellent starting point.  An
in-depth look at the original oblique decision tree algorithm used to
generate the trees for our experiments can be found in a paper by
Murthy et al. \shortcite{mu3}.

\subsection{Neural Networks}

An artificial neural network consists of nodes, roughly analogous to
human neurons, arranged in a series of layers.  All the nodes of each
layer can be fully connected to the nodes in the next.  Weights
between nodes indicate how a series of inputs are to be transformed
into output; for example, how attributes describing an object (input)
determine the object's classification (output).  ``Hidden nodes''
occur between the input layer of nodes and the output layer.  Node
weights are updated via a hill-climbing error-minimization procedure.
In other words, as the neural net learns, weights are updated such
that the overall accuracy of the neural net improves.

	The error-minimization procedure implemented by SLSS is known
as back-propagation.  Back-propagation modifies weights from the
output nodes backwards to the nodes accepting input.  As mentioned,
weights are only altered when back-propagation improves the neural
net's overall accuracy.  Of course, to determine overall accuracy,
some kind of pre-classified data is required (discussed below).

The ANN implemented by SLSS contains 13 input nodes, 13 hidden nodes,
and 5 output nodes.  The 5 output nodes correspond to the five target
classifications the neural net has been trained to produce.  A
classification is based on the output node with the largest value.
Following a crude Hubble sequence, the five classes chosen by SLSS
are: E, S0, Sa+Sb, Sc+Sd, and Irr.

        Because the ANN developed by SLSS is a supervised neural
network, the ANN must first be trained on some pre-classified set of
data.  Subsets of this data were reserved and used to test the
accuracy of the neural net.  Because the ANN must first be trained,
its accuracy is wholly dependent on the accuracy of the training data.
Good training data is essential to the production of a good
classifier.

\begin{table}
\caption{SLSS results with an ANN.  Overall accuracy = 64.1\%.}
\begin{tabular}{lrrrrr}
{\bf Class}	&E	&S0	&Sa+Sb	&Sc+Sd	&Irr	\\
E       &203    &77     &25     &1      &5      \\
S0      &109    &229    &240    &7      &2      \\
Sa+Sb   &12     &85     &1281   &218    &15     \\
Sc+Sd   &1      &4      &304    &415    &36     \\
Irr     &0      &0      &53     &69     &126    \\
\end{tabular}
\end{table}

\begin{table}
\caption{Lauberts \& Valentijn's results with their own automated
classifier.  Overall accuracy = 56.3\%.}
\begin{tabular}{lrrrrr}
{\bf Class}     &E      &S0     &Sa+Sb  &Sc+Sd  &Irr	\\
E       &197    &87     &17     &5      &5      \\
S0      &184    &218    &155    &28     &2      \\
Sa+Sb   &106    &12     &791    &664    &38     \\
Sc+Sd   &22     &11     &24     &631    &72     \\
Irr     &22     &9      &31     &42     &144    \\
\end{tabular}
\end{table}

The results obtained by the SLSS neural net are given in table 1.

\subsection{Decision Trees}

A decision tree can be thought of as the outline of a decision
process.  As with any tree-like data structure, a decision tree
consists of both internal and leaf nodes.  Internal nodes correspond
to choices to be made from the set of training data; leaf nodes
correspond to conclusions.  A path from the root node to a specific
leaf constitutes a decision.

        Throughout the rest of this paper, we will be concerned only
with binary decision trees.  Binary decision trees can make only
yes/no decisions at each internal node.

        To determine the classification of a specific object, the
attribute describing the objects are passed through the decision tree
starting at the root.  Each internal node might contain a test of the
form $a_{i}X > k$, where $a_{i}$ is the $i$th attribute for any example
X, and $k$ is a test value.  This defines a one-dimensional hyperplane
across attribute $i$.

A decision is based on whether a specific attribute $a_{i}$ for a
given example is greater or less than a value $k$ stored in the tree.
If the example's attribute is greater, then the ``yes'' branch is
followed, otherwise the opposite ``no'' branch will be taken.  This
process continues recursively until a leaf node (a conclusion) is
reached.  Tests of this form -- tests that classify objects by
dividing a data space -- are known as {\bf splits}.  Splits that
segment a space only along a single dimension (with a single
attribute) such as the above are commonly referred to as {\bf
axis-parallel} splits.

        Another test that could be performed at each node is an
oblique split:

$${\sum_{i=1}^{d} (a_{i}X_{i}) + a_{d+1} > 0}$$

\noindent where each example $X$ has $d$ attributes.  Here, decisions
are made by determining whether an object lies above or below a
$d$-dimensional hyperplane defined by each attribute $a_{i}$.  As with
axis-parallel tests, whether or not an example lies above the
hyperplane indicates whether the ``yes'' or ``no'' branch will be
followed.  Splits of this form are referred to as {\bf oblique}.

\medskip

        The particular trees we have grown for our experiments contain
almost exclusively oblique splits and will be called oblique
decision trees.  Oblique decision trees are known to require, in
general, far fewer nodes to accurately describe data.  This is because
oblique trees take advantage of Occam's razor.  Intuitively, given two
trees, each equally accurate on a set of training data, the decision
tree with the fewer number of nodes will be expected, in general, to
make more accurate decisions on new, previously unseen data.  The
algorithm we use to construct our decision trees is an extension of
Murthy et al.'s Oblique Classifier 1 (OC1) algorithm.

\section{Training Oblique Trees with OC1}

Like the SLSS ANN, OC1 learns how to classify a data set by first {\em
training} on a pre-classified subset of the data.  The decision tree
grown from this training set can then be used to classify the
remainder of the data as well as new examples.

As described above, OC1 uses both oblique and axis-parallel
hyperplanes to partition sets of data.  The internal nodes of a tree
generated by OC1 therefore constitute either axis-parallel or oblique
splits.  OC1 searches through the data to find the best split for each
node.  The quality of a split is determined by a measure known as {\bf
impurity}.  Impurity is a heuristic measure of how {\em poorly} a
certain split will separate data.  The goal of OC1 is to find splits
which minimize the overall impurity of a decision tree.

        While most of the searching that OC1 performs is local
deterministic hill-climbing (although, as we are using impurity
measures, it would be more accurate to say ``hill descending''), some
randomization has been introduced to determine placement of the
initial hyperplane and to escape from local minima.  This stochastic component
of the algorithm is
necessary because of the enormous size of the search space: Given $n$
objects with $d$ dimensions (attributes), the number of distinct
oblique hyperplanes that can separate the $n$ objects is $2^d \cdot {n
\choose d}$.  This is a much greater value than the $n
\cdot d$ distinct axis-parallel splits that can separate $n$ objects.
By performing multiple local searches in this manner, OC1 can come
very close to optimal solutions without the overhead of an exhaustive
search.  In fact, as Heath has shown, the problem of finding an
optimal oblique split is NP-Complete \cite{he}.  In other words, it
is doubtful that any algorithm could find an optimal oblique split in
an amount of time that is a polynomial function of $n$ and $d$: such
an algorithm would more than likely require an exponential amount of
processing time.
       OC1 takes the best split it can find for an internal node
before moving recursively to the next.  Thus, OC1 performs a {\em
greedy} search for each split.  OC1 does not use any form of
``lookahead'' to determine whether potentially bad splits might result
in good trees.  Murthy and Salzberg \shortcite{mu1} have shown that
such a mechanism provides only marginal, if any, improvement.  Like
OC1, most decision tree algorithms restrict their search to the space
of data, rather than searching unnecessarily through the space of
trees.

	Heath \shortcite{he} points out that the stochastic element of algorithms such
as OC1 can be advantageous.  By generating multiple classifiers from a set of
data, classification of new data can be determined by popular vote.  That is,
the most common classification of an object among multiple classifiers
determines the object's overall class; thereby reducing the chance of
classification error.  Thus, OC1 could be used to generate multiple decision
trees, a decision forest, from a set of training data.  The classification of a
new object would be determined by the most common classification to occur in
the decision forest.

\subsection{Impurity Measure}

The impurity measure we have chosen to use with OC1 is known as the
``twoing'' criterion and can be defined as follows (Breiman et
al. 1984, Salzberg et al. 1994):

$${(p_{L} \cdot p_{R}) (\sum_{j} |(p(j|L) - p(j|R))|)^2}$$

\noindent where $p_{L}$ and $p_{R}$ are, respectively, the proportion
of examples on the left and right side of a split, and both $p(j|L)$
and $p(j|R)$ represent the proportions of class $j$ on the left and
right sides of the split.  This criterion assigns higher values to
hyperplanes that come close to splitting the data in half and to
hyperplanes that split cleanly between classes.  When examples from
the same class are split apart, the values returned by the twoing
criterion indicate increased impurity.  As mentioned above, OC1
strives to minimize this measure of impurity for each split.

\section{Experiments with Decision Trees}

Here we repeat the experiments performed by SLSS with the decision
tree generating algorithm, OC1.  We also include some experiments of
our own.  As in the SLSS experiments, data is taken from the ESO-LV
catalog \cite{lau}.  The 13 parameters (attributes) used to describe
each object in their experiments are given below (taken from SLSS
1992):

\def\bhang{\smallskip\noindent\hang $\bullet\ $}

\medskip\bhang $<B - R>$: average color in region with $B$ surface brightness
  20.5 to 26.

\bhang $N_{oct}^{B}$: exponent of the fit of a generalized de
Vaucouleurs law to [the galaxy profile] $B$ octants (N = 0.25
corresponds to a perfect elliptical galaxy and N = 1 to a pure
exponential disk).

\bhang $log(D_{80}^{B}/D_{e}^{B})$, where $D_{80}^{B}$ and $D_{e}^{B}$ are
the major diameters of the ellipses at 80 per cent and half total $B$
light respectively.

\bhang $\nabla_{rad}^{tan}$: arctangent of the absolute value of the ratio of
the mean tangential and radial gradients, which is an indicator of the
degree of asymmetry of the galaxy image.

\bhang $\mu_{oct}^{B}$: $B$ central surface brightness from the fit of a
generalized de Vaucouleurs law to $B$ octants.

\bhang $log(b/a)$, where $b/a$ is the galaxy axial ratio.

\bhang $E_{err}^{fit}$: error in ellipse fit to $B$ isophotes at $B$ surface
brightness 23.

\bhang $\nabla_{R_{e}}$: gradient of the $B$ surface brightness profile at
$D_{e}^{B}$.

\bhang $log(D_{26}^{B}/D_{e}^{B})$, where $D_{26}^{B}$ is the major diameter
of the ellipse at 26 $B$ mag $arcsec^{-2}$.

\bhang $N_{oct}^{R}$: exponent of the fit of a generalized de Vaucouleurs law
to
R octants.

\bhang $\mu_{0}^{B}$: average $B$ surface brightness within 10 arcsec
diameter circular aperture.

\bhang $\mu_{e}^{B}$: $B$ surface brightness at half total $B$ light.

\bhang $\mu_{e}^{R}$: $R$ surface brightness at half total $R$ light.

\bigskip

Only those galaxies with ESO visual diameter $\ge 1$ arcmin and at
high Galactic latitude ($|b| > 30\deg$) were considered.  All galaxies
have been morphologically classified via visual examination.
According to SLSS, these attributes were chosen
because they are distance independent and because they are very
similar to those used by Lauberts \& Valentijn to perform their own
automated classification (results of which are presented in table 2).

The final data set contains 5217 galaxies.  SLSS randomly sorted this
data into two sets of 1700 and 3517 objects to be used, respectively,
for training and testing.  We trained our decision trees using both
this method and with a five-fold cross-validation experiment.  The
five major classes, determined by Lauberts \& Valentijn, were
binned based on the following criteria: {\bf E} ($-5.0 \le T < -2.5$,
466 galaxies), {\bf S0} ($-2.5 \le T < 0.5$, 851 galaxies), {\bf
Sa+Sb} ($0.5 \le T < 4.5$, 2403 galaxies), {\bf Sc+Sd} ($4.5 \le T <
8.5$, 1132 galaxies), and {\bf Irr} ($8.5 \le T \le 10.0$, 365
galaxies), where $T$ is an object's type.

	Tables 1 and 2 compare the performance of the SLSS ANN to the
Lauberts \& Valentijn classifier.  Rows in these tables reveal visual
type distributions; columns depict automated type distribution.  Left
to right diagonals are the values for which both human and automated
classifiers perfectly agree.  Overall accuracy for each of the
classifiers is given in the caption above each table.  Notice that the
ANN produces superior performance.

	Table 3 outlines the overall performance of OC1 using a
five-fold cross-validation experiment.  Cross-validation studies of
this sort are known to produce reliable estimates of accuracy while
avoiding ``optimistic'' bias \cite{sa2}.  The ``leaves'' column
denotes the number of possible decisions contained within each tree.
To perform five-fold cross-validation, all 5217 galaxies were split
into five equal-sized sets.  4/5 of this data was reserved for
training and the remaining 1/5 for testing.  The process was repeated
another four times, thus allowing each set to be used once as a test
set.

\begin{table}
\caption{Five-fold cross-validation on all 5217 objects.  {\em Right:}
Modest random search (5 random perturbations, 20 restarts); {\em
Left:} Extensive random search (100 random perturbations; 100 random
restarts).}
\begin{tabular}{rrrlrr}
	&\multicolumn{2}{l}{Modest Search}	&\multicolumn{2}{l}{Extensive Search}	\\
Fold	&Accuracy	&Leaves	&Accuracy	&Leaves	\\
1	&64.5	&11	&64.8   &15    	\\
2	&64.4	&7	&65.3   &9     	\\
3	&62.3   &8	&62.9   &8     	\\
4	&64.1   &9	&65.3   &8 	\\
5	&63.8   &9	&64.7   &10	\\
Average	&63.8	&8.8	&64.6	&10.0	\\
\end{tabular}
\end{table}

        The left half of table 3 demonstrates OC1's performance with a
modest amount of random search; the right half illustrates the
performance of a more exhaustive search.  From this table it is clear
that additional search result in only a marginal accuracy improvement
(on average, less than 1\%).  Interestingly, the improvement in
accuracy seems to require an increase in the number of leaf nodes
(decisions) contained within each tree.  Because additional search
requires more processing time, we use only the default, modest degree
of search in the remaining experiments.

\begin{table}
\caption{Results obtained with AP-OC1.  Overall accuracy = 63\%.
Number of leaves = 43.}
\begin{tabular}{lrrrrr}
{\bf Class}     &E      &S0     &Sa+Sb  &Sc+Sd  &Irr	\\
E       &207    &75     &23     &2      &4      \\
S0      &101    &227    &250    &7      &2      \\
Sa+Sb   &5      &78     &1336   &175    &17     \\
Sc+Sd   &0      &3      &390    &302    &65     \\
Irr     &0      &0      &57     &51     &140    \\
\end{tabular}
\end{table}

\begin{table}
\caption{Results obtained with the most
accurate tree produced by OC1.  Overall accuracy = 63\%.  Number of
leaves = 23.  5 random perturbations; 20 random restarts.}
\begin{tabular}{lrrrrr}
{\bf Class}     &E      &S0     &Sa+Sb  &Sc+Sd  &Irr	\\
E       &188    &102    &17     &1      &3 	\\
S0      &84     &259    &236    &7      &1  	\\
Sa+Sb   &3      &100    &1279   &220    &9  	\\
Sc+Sd   &0      &1      &335    &375    &49 	\\
Irr     &0      &0      &49     &80     &119	\\
\end{tabular}
\end{table}

        The results in table 4 were obtained by running OC1
exclusively in axis-parallel mode (AP-OC1), which requires no random
search.  Like SLSS, only 1700 randomly chosen objects were used for
training.  Cross-validation was not used so that our experiments would
remain consistent with those performed by SLSS.  Notice that even with
this very simple method of constructing decision trees, we are but one
percent away from the overall accuracy obtained by the SLSS neural net
(table 1).

\begin{figure}
\epsffile[0 0 233 144]{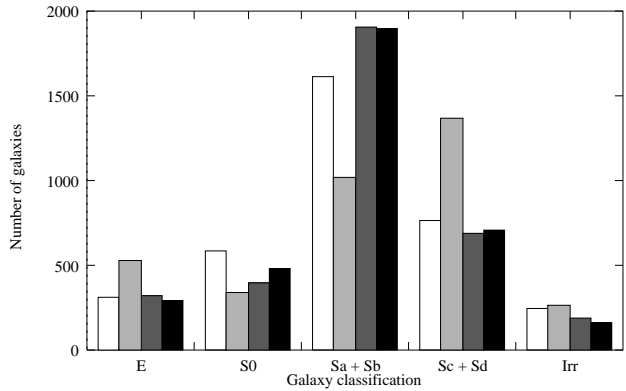}
\caption{Distribution of classification for the
test sample (3517 galaxies).  White bars indicate visual
classification, light gray indicate Lauberts \& Valentijn's automated
classifier, dark gray is the SLSS ANN; black represents OC1.}
\end{figure}

        Table 5 displays results of the most accurate tree taken from
five runs of OC1 with different random seeds.  Again only 1700
randomly chosen objects were used for training.  Notice that the
incorporation of oblique splits has resulted in a tree that is half
the size of the the axis-parallel tree.  Still, this tree is large for
an oblique decision tree.  Most trees produced by OC1 are much
smaller.  In one case, for example, by using a different information
measure known as the gini index, an oblique tree composed of 6 leaves
with 62\% accuracy was constructed.  The gini index, which is a
measure of the probability of misclassification over a set of
instances, has been modified for use as an impurity measure in the OC1
package (Murthy et al. 1994, Breiman et al. 1984).  Different methods
for evaluating the quality of a split, including the gini index and
the twoing criterion, extend the power and flexibility of OC1.

        The five randomized OC1 runs produced trees with an average of
17 leaves and an average accuracy of 61\%.  Apparently the reduction
in training examples has had an adverse affect on the quality of our
decision trees.  Furthermore, while normally we would expect the
smaller trees to be more accurate classifiers, the fact that one of
the larger trees achieves the highest accuracy indicates that the data
is complex and difficult to generalize.  Our result from table 3, that
additional accuracy seems correlated to additional leaves, also
supports this conclusion.

\medskip

        Figure 1 compares classification distributions from each of
the classifiers to the original visual distribution.  Notice how
closely related the neural net and decision tree distributions are,
even though none of the output from any of the automated classifiers
parallels the visual classifications with considerable accuracy.
While Lauberts \& Valentijn's ESO classifier seems to reverse the
distribution of Sa+Sb and Sc+Sd galaxies found in the visual
classification, both the ANN and decision tree reflect this pattern.

\section{Discussion}

We have shown that decision trees can be used to determine the
morphological classification of galaxies with reasonable success.
Furthermore, our comparisons of a neural net classifier to that of a
decision tree algorithm have produced similar results.  Errors made by
both classifiers can be attributed to one or more of the following
problems: 1) there are errors in the visual classification of the 5217
galaxies which comprise both the training and test sets, or 2) none of
the attributes used to describe the data provides sufficient
information for accurate classification.  SLSS account for this by
noting that the classifications are based on plate material rather
than CCD frames, and that the parameters used to describe the
galaxies were chosen somewhat arbitrarily.

	As can be seen in both the ANN results and those obtained by
our decision trees (tables 1, 4, and 5), while non-neighbor classes
can, potentially, be easily separated, neighbor classes cannot.  In
other words, while trees grown to discriminate E-type galaxies from
Sa+Sb-types might typically be very accurate, trees that distinguish
between neighboring types such as E and S0 would have very poor
accuracies.  In fact, as SLSS noticed, scoring accuracy in terms of
nearest neighbor classifications results in roughly a 90\% accuracy.

	Table 6 demonstrates that multiple decision trees can, in
fact, be generated to easily distinguish between different regions
along the continuum of classifications.  The decision trees used to
produced these results were trained on the small 1700 object set used
above.  Clearly, extremely accurate and simple trees can be induced
from this simplified data.  With these trees, galaxies can now be
confidently classified to larger, overlapping regions.  For example,
by using a majority vote among all six trees, a galaxy might be
classified either to the E-S0 region or to the S0-Sa+Sb-Sc+Sd region.
In one experiment, the trees in table 6 were manually assembled to
produce an E-Sa+Sb-Irr classifier with a 90.7\% accuracy.

\begin{table}
\caption{Accuracy and tree size (in number of leaves) of non-neighbor
decision trees.}
\begin{tabular}{lccclcc}
{\bf Tree}	&Acc	&Lvs	&&{\bf Tree}	&Acc &Lvs	\\
1. E / Sa+Sb	&96.4	&3	&&4. S0 / Sc+Sd	&91.4   &2      \\
2. E / Sc+Sd	&97.3	&2	&&5. S0 / Irr	&95.7   &2      \\
3. E / Irr	&95.7	&2	&&6. Sa+Sb / Irr&92.8	&3	\\
\end{tabular}
\end{table}

In one last attempt to overcome the five-classification ``fuzziness''
of the data, we tried growing two new trees: one to separate E and
Irr-type galaxies from spirals, and another to identify the S0, Sa+Sb,
and Sc+Sd-types in the spiral subset.  The result was a modest
increase in accuracy to 66\%.  While this result could most likely
have been achieved by increasing the number of random searches
performed by OC1, by initially filtering out the spirals, we were able
to direct the search in a direction we wanted to explore.  By reducing
the search space in this manner, we also reduced processing time.

        Finally, even though neither the decision tree nor the ANN
produced remarkable results, global classifications by the two differ
for less than 3\% of the test set.  (Attempts have not yet been made
to determine accuracy by comparing classifications example by
example.)  Furthermore, misclassifications made by the oblique
decision tree in table 5 match roughly 83\% of the misclassifications
made by the SLSS ANN.  The similar results obtained by these two very
different classifiers does, perhaps, point to the existence of error
in the original data.  At the very least, our results confirm the
discovery made by the SLSS ANN that the distinction between
neighboring classes appears to be poorly defined.  The two
classification algorithms may ultimately be discovering a more
accurate way by which to classify the original Lauberts \& Valentijn
data.

\section*{Software}

The OC1 software discussed in this paper is available over the Internet
via anonymous ftp.  The package contains full source code and extensive
documentation.  For instructions on how to retrieve the package, the
authors request inquiries to be made to either salzberg@cs.jhu.edu or
murthy@cs.jhu.edu.

\section*{Acknowledgments}

Thanks to Avi Naim for providing the SLSS attribute choices for
the Lauberts \& Valentijn data, as well as some helpful comments.
Thanks to Steven Salzberg and Sreerama K. Murthy for providing the OC1
decision tree package, as well as suggestions towards its proper use.  Thanks
to Chris League for revision advice.
This work has been supported by the Medium Deep Survey which is funded
by STScI grants GO 2684.0X.87A and GO 2684.0X.91A.

\bsp

\label{lastpage}

\end{document}